\documentclass[lettersize,journal]{IEEEtran}
\usepackage{amsmath,amsfonts}
\usepackage{algorithmic}
\usepackage{algorithm}
\usepackage{array}
\usepackage[caption=false,font=normalsize,labelfont=sf,textfont=sf]{subfig}
\usepackage{textcomp}
\usepackage{stfloats}
\usepackage{url}
\usepackage{verbatim}
\usepackage{graphicx}
\usepackage{cite}
\usepackage{color}
\usepackage{multirow}
\usepackage{booktabs}
\usepackage{hyperref}

\begin{document}

\title{Multi-Scale Accent Modeling and Disentangling \\ for Multi-Speaker Multi-Accent \\ Text-to-Speech Synthesis}

\author{Xuehao Zhou,~\IEEEmembership{Student Member,~IEEE,}
        Mingyang Zhang,~\IEEEmembership{Member,~IEEE,}      
        Yi Zhou,~\IEEEmembership{Member,~IEEE,}             \\
        Zhizheng Wu,~\IEEEmembership{Senior Member,~IEEE,}
        and Haizhou Li,~\IEEEmembership{Fellow,~IEEE}
\thanks{Xuehao Zhou and Yi Zhou are with the Department of Electrical and Computer Engineering, National University of Singapore, Singapore (e-mail: xuehao.zhou@u.nus.edu; yi.zhou@u.nus.edu)}
\thanks{Mingyang Zhang is with Hong Kong Generative AI Research and Development Centre, The Hong Kong University of Science and Technology, Hong Kong SAR, China. (email: mileszhang@ust.hk)}
\thanks{Zhizheng Wu and Haizhou Li are with the Shenzhen Research Institute of Big Data, School of Data Science, The Chinese University of Hong Kong, Shenzhen, China. Haizhou Li is also with the Department of Electrical and Computer Engineering, National University of Singapore, Singapore (email: wuzhizheng@cuhk.edu.cn; haizhouli@cuhk.edu.cn)}}


\maketitle

\begin{abstract}
Generating speech across different accents while preserving speaker identity is crucial for various real-world applications. However, accurately and independently modeling both speaker and accent characteristics in text-to-speech (TTS) systems is challenging due to the complex variations of accents and the inherent entanglement between speaker and accent identities. 
In this paper, we propose a novel approach for multi-speaker multi-accent TTS synthesis that aims to synthesize speech for multiple speakers, each with various accents. Our approach employs a multi-scale accent modeling strategy to address accent variations on different levels. Specifically, we introduce both global (utterance level) and local (phoneme level) accent modeling to capture overall accent characteristics within an utterance and fine-grained accent variations across phonemes, respectively. 
To enable independent control of speakers and accents, we use the speaker embedding to represent speaker identity and achieve speaker-independent accent control through speaker disentanglement within the multi-scale accent modeling.  
Additionally, we present a local accent prediction model that enables our system to generate accented speech directly from phoneme inputs. 
We conduct extensive experiments on an English accented speech corpus.
Experimental results demonstrate that our proposed system outperforms baseline systems in terms of speech quality and accent rendering for generating multi-speaker multi-accent speech.
Ablation studies further validate the effectiveness of different components in our proposed system. 
\end{abstract}


\section{Introduction}
\label{sec:introduction}
\IEEEPARstart{T}{EXT}-TO-SPEECH (TTS) systems play a crucial role in human-computer interaction, converting raw text into natural-sounding speech. Over the years, TTS technology has made significant progress, from statistical parametric modeling \cite{zen2009statistical, ze2013statistical} to end-to-end (E2E) architectures \cite{wang2017tacotron, shen2018natural, li2019neural, ren2020fastspeech, elias2021parallel, kim2021conditional, wang2023neural}, which generate high-quality and human-like speech directly from text inputs.
While a standard TTS system generates speech with a single speaker's voice, recent research focuses on building multi-speaker TTS systems \cite{chen2020multispeech, fan2015multi, casanova2022yourtts, gong2024zmm} that generate speech for multiple speakers, providing diverse speech outputs for personalized applications, such as voice cloning.
However, for cross-regional applications, 
there is an increasing demand for speech outputs that combine speaker diversity with a wide range of accent expressions.
For example, an audiobook that supports multiple accents can enhance the user experience for listeners from different regions, and a language learning platform with various accents helps bridge understanding gaps for learners with diverse linguistic backgrounds.
To meet these demands, it is crucial to develop multi-speaker multi-accent TTS systems that can generate voices of multiple speakers, each with various accents.

An ideal approach to develop a multi-speaker multi-accent TTS system is to train the model on multi-speaker multi-accent speech corpus, where each speaker's recordings include multiple accents. However, such comprehensive datasets are often unavailable, as each speaker is typically associated with only one accent tied to their native region.
To address this limitation, this paper investigates to develop a generalized multi-speaker multi-accent TTS system using an existing multi-speaker dataset from diverse regions, where each speaker has a single accent. 
To enable flexible combinations of speakers with different accents for multi-speaker multi-accent speech synthesis, 
independent modeling of speaker and accent characterisitcs is required. 
To model speaker identity, we use the widely adopted technique of speaker embedding \cite{jia2018transfer, yang2021ganspeech, wang2024usat}, which learns speaker-discriminative representations from large-scale speaker datasets trained on a speaker classification task \cite{wan2018generalized}. The speaker embedding, when integrated into multi-speaker TTS systems, effectively controls voices of multiple speakers. Building on this, we extend multi-speaker TTS to multi-speaker multi-accent TTS by exploring accurate and independent accent modeling.

The perception of foreign accents in second language speakers is significantly influenced by both phonetic \cite{reinisch2014lexically, loots2011automatic} and prosodic \cite{cho2005prosodic, mareuil2006contribution, vaissiere2004identifying} variations, which complicate accent modeling.
Among these factors, phoneme level speech information plays an important role in accurately capturing accent representations \cite{angkititrakul2006advances}.
For example, vowel formants differ across accents \cite{watson1998acoustic, hansen1995foreign}, and phoneme level pitch variations have a greater effect on accent perception than utterance level variations \cite{ding2016prosodic}.   
Furthermore, accent representations vary within an utterance \cite{terken1991fundamental, fletcher2005intonational}. 
Studies indicate that pitch patterns exhibit different ranges across phonemes \cite{yan2003analysis, bolinger1958theory}, 
and vowel durations show notably greater differences than consonant durations \cite{yan2003analysis}.
These phoneme level elements are essential for capturing fine-grained variations of accented speech, distinguishing accent modeling from typical style modeling \cite{wang2018style, zhang2019learning} and speaker modeling approaches, which primarily focus on learning global representations. 
Another challenge for accent modeling is the inherent entanglement between speaker and accent identities, particularly when generating speech with unseen accents for target speakers. This entanglement can interfere with the voices of target speakers, resulting in degraded speaker similarity in the generated speech. 
In this paper, we propose a multi-scale accent modeling and disentangling approach for multi-speaker multi-accent TTS synthesis. 
First, our method employs both global and local accent modeling to comprehensively address the complex variations of accents. 
Global accent modeling provides an utterance level representation of accented speech, capturing high level accent characteristics related to fundamental phonological features and overall prosodic patterns. However, global modeling alone may lack details of accent fluctuations, such as phoneme pronunciation patterns and segmental prosodic differences, potentially leading to weak accent expression in the generated speech. 
To address this limitation, we introduce local accent modeling to capture fine-grained accent representations on the phoneme level, 
including stress, intonation, and duration patterns, which are crucial for representing accent variations within an utterance. Local accent modeling complements global accent modeling, providing comprehensive and accurate descriptions of accent characteristics.

Second, for flexible multi-speaker multi-accent speech synthesis, speaker-independent accent modeling is essential. 
To achieve this, we perform speaker disentanglement within both global and local accent modeling, capturing accent characteristics at both scales in a speaker-independent manner. 
As a result, our method enables accurate and independent control of accents in the generated speech.
Third, while local accent modeling produces phoneme level accent representations from the Mel-spectrogram, it heavily relies on reference speech containing the target accent during inference. Moreover, the reference speech must align with the TTS input content, and force-alignment is required. 
Inspired by studies on accent recognition showing that accent variations are closely associated with the phonetic realization of speech \cite{zhang2021accent, kat1999fast, biadsy2011dialect}, 
we propose a local accent prediction model that predicts phoneme level accent representations directly from phoneme inputs, thereby eliminating the need for reference recordings during inference. This enhances the practical applicability of our TTS system across diverse contexts. 
The contributions of this paper are summarized as follows:
\begin{itemize}
    \item We propose a multi-scale accent modeling approach that produce both global and local accent representations, comprehensively capturing accent variations to achieve accurate accent rendering.
    \item We perform speaker disentanglement to achieve speaker-independent accent modeling, enabling the independent control of accents in the generated speech. 
    \item We introduce a local accent prediction model that enables our TTS system to generate multi-accent speech directly from phoneme inputs, without requiring reference accented speech during inference. 
\end{itemize}

The rest of this paper is organized as follows: Section \ref{sec:related_work} reviews related studies and summarizes the research gap for multi-speaker multi-accent TTS synthesis. Section \ref{sec:method} introduces our proposed approach and system architecture. We present the experimental setup and results in Section \ref{sec:exp_setup} and \ref{sec:exp_results}, respectively. Section \ref{sec:conclusion} concludes this paper and discusses limitations of this study with future work.

\section{Related Work}
\label{sec:related_work}
In this section, we first review related work on expressive TTS, as it shares some similarities with accented TTS. We then review studies specifically focusing on accented TTS.

\subsection{Expressive TTS}
Extensive research has explored developing expressive TTS systems that generate speech with the target style.
One typical method, global style token (GST) \cite{wang2018style}, learns a style embedding from reference speech without the need for style labels. This approach has been shown to effectively convey stylistic information in TTS systems \cite{valle2020mellotron, neekhara2021expressive, nishihara2024low}. 
Another unsupervised technique, variational autoencoder (VAE) \cite{zhang2019learning}, encodes style representations from expressive speech into a latent space that can be manipulated for style control. VAE-based methods have been widely used for style transfer \cite{guan2023interpretable} across intra-speaker, inter-speaker, and unseen speaker scenarios \cite{xue2021cycle}, as well as for style enhancement \cite{chen2024stylespeech, li2024cross}. 
In multi-speaker expressive TTS, generating speech with an unseen style to the target speaker often suffers from performance degradation due to the entanglement of style and speaker attributes.
To address this problem, style disentanglement approaches have been investigated.
Studies show that adversarial training is an effective technique for learning disentangled style or emotion and speaker representations \cite{whitehill2019multi, li2022cross, dutta2024zero}.  
Additionally, disentangling language identity for multi-lingual multi-speaker expressive TTS has also been studied \cite{choi2024mels}.

The aforementioned studies have made substantial progress in expressive TTS. 
While both expressive speech and accented speech exhibit variations on the utterance level, accent speech also involves significant variations in segmental speech units, which are crucial for accurate accent modeling, as discussed in Section \ref{sec:introduction}. 
Although fine-grained emotion modeling has been explored in \cite{lei2022msemotts, tang2024ed}, these methods focus on emotion rather than accent. 
Inspired by advancements in expressive TTS, this paper focuses on effective accent modeling for multi-speaker multi-accent TTS synthesis. 

\subsection{Accented TTS}
Accented TTS systems aim to generate speech with the target accent from text inputs. 
Zhou et al. \cite{zhou2024accented} propose an accented TTS framework consisting of an accented front-end and an accented acoustic model with additional pitch and duration predictors, addressing phonetic and prosodic variations of accented speech. 
Zhang et al. \cite{zhang2023zero} introduce a residual layer appended to the encoder to learn accented phoneme representations by mapping native speech to accented speech. 
However, both methods focus on fine-tuning a single accent with limited data. 
Tinchev et al. \cite{tinchev2023modelling} present a data augmentation approach for accent modeling. They use voice conversion to augment the target accent data, and then build a multi-speaker multi-accent TTS system with both real and synthetic data.
Nguyen et al. \cite{nguyen2023syntacc} propose a multi-accent TTS framework utilizing a weight factorization approach. They decompose each weight matrix of the letter-to-sound component into shared and accent-dependent factors. 
Zhang et al. \cite{zhang2023towards} develop a multi-accent TTS system that controls accents in the encoder, which is trained on an auxiliary accent classification task to generate multi-accent phoneme representations.

There have also been efforts to extract accent representations from accented speech. 
Multi-level VAE \cite{melechovsky2024accent, melechovsky2024dart} has been studied to model both accent and speaker representations in accented TTS systems.
Zhong et al. \cite{zhong2024accentbox} introduce a two-stage training pipeline for zero-shot accent generation. They first train a speaker-independent accent encoder and then build an accented TTS system conditioned on the pre-trained accent encoder. 
However, these methods primarily capture accent characteristics at the global scale. 
Ma et al. \cite{ma2023accent} propose leveraging bottleneck features from an automatic speech recognition (ASR) model for accent transfer. While this method enables fine-grained accent modifications, it heavily depends on a well-trained ASR model.

\begin{figure*}
\centering
\includegraphics[width=0.95\textwidth]{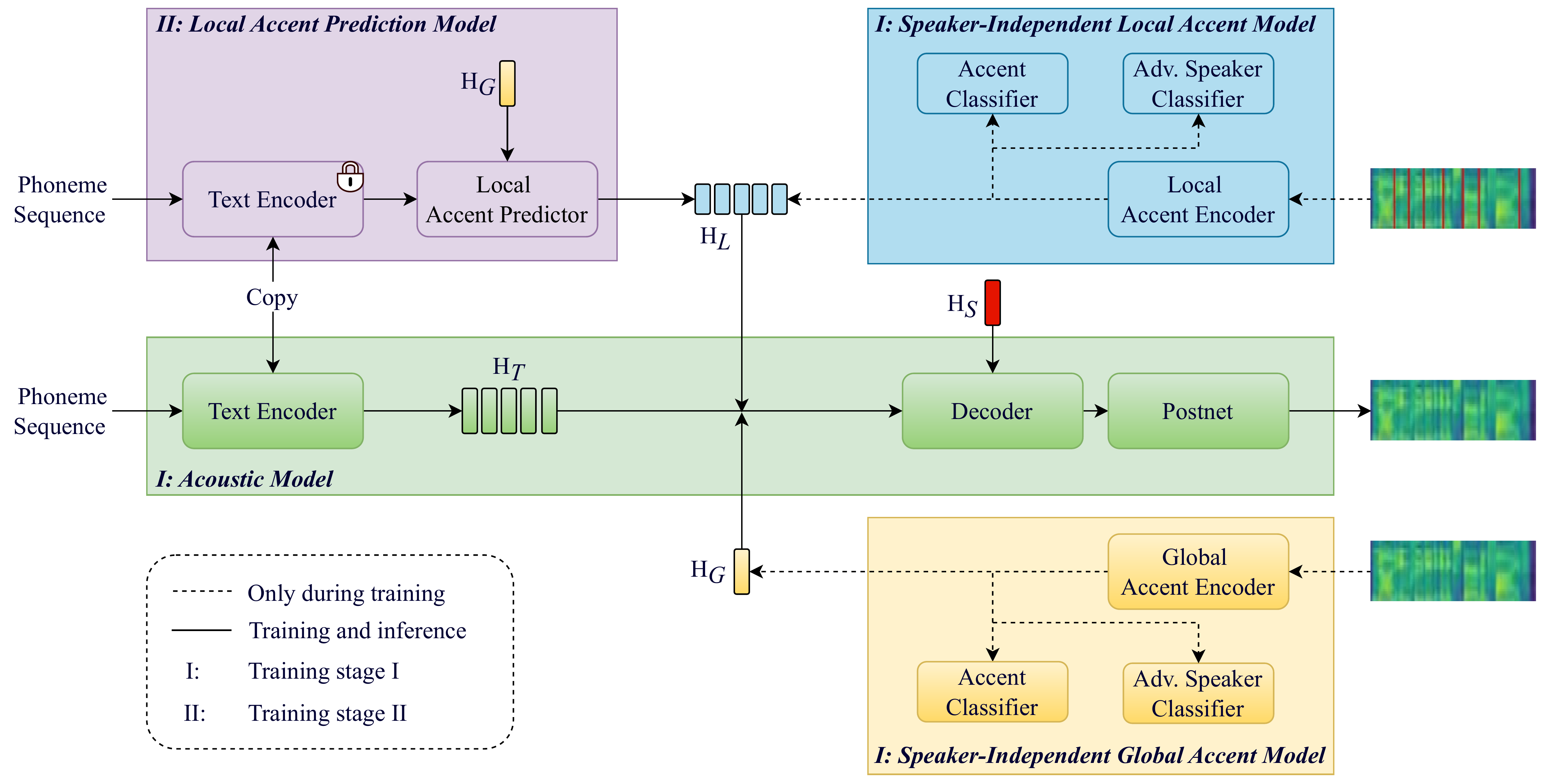}
\caption{The architecture of the proposed multi-speaker multi-accent TTS framework in two training stages. The first stage is to train the acoustic model with the speaker-independent global and local accent models, and the second stage is to train the local accent prediction model. The speaker embedding $H_{S}$ is extracted from a pre-trained speaker encoder. Speech waveforms are generated by a pre-trained neural vocoder from the predicted Mel-spectrogram.}
\label{fig_sys}
\end{figure*}

In summary, fine-grained and independent accent modeling within TTS systems for multi-speaker multi-accent speech synthesis remains underexplored. 
Liu et al. \cite{liu2024controllable} propose a method to control accent intensity on both coarse and fine-grained levels, but they primarily focus on accent intensity rather than multi-speaker multi-accent speech synthesis. 
Motivated to address these gaps, this paper investigates a multi-scale accent modeling approach jointly optimized with the TTS model to achieve accurate and effective accent rendering.
Additionally, we extend to multi-speaker multi-accent TTS synthesis by investigating the disentanglement between accents and speakers.

\section{Methodology}
\label{sec:method}
Our proposed multi-speaker multi-accent TTS framework with multi-scale accent modeling and disentangling approach is illustrated in Fig. \ref{fig_sys}.
The acoustic model (AM) predicts the Mel-spectrogram from the phoneme sequence, serving as the backbone of the TTS framework.
The speaker-independent global accent model (SIGAM) and speaker-independent local accent model (SILAM) produce utterance and phoneme levels accent representations, respectively, while the local accent prediction model (LAPM) predicts phoneme level accent representations.
Accented speech is generated by a pre-trained neural vocoder from the predicted Mel-spectrogram.

In this section, we introduce the AM, SIGAM, SILAM, and LAPM, along with their respective objective functions. We also describe the training and inference stages of our proposed framework. 

\begin{figure*}
\centering
\includegraphics[width=0.8\textwidth]{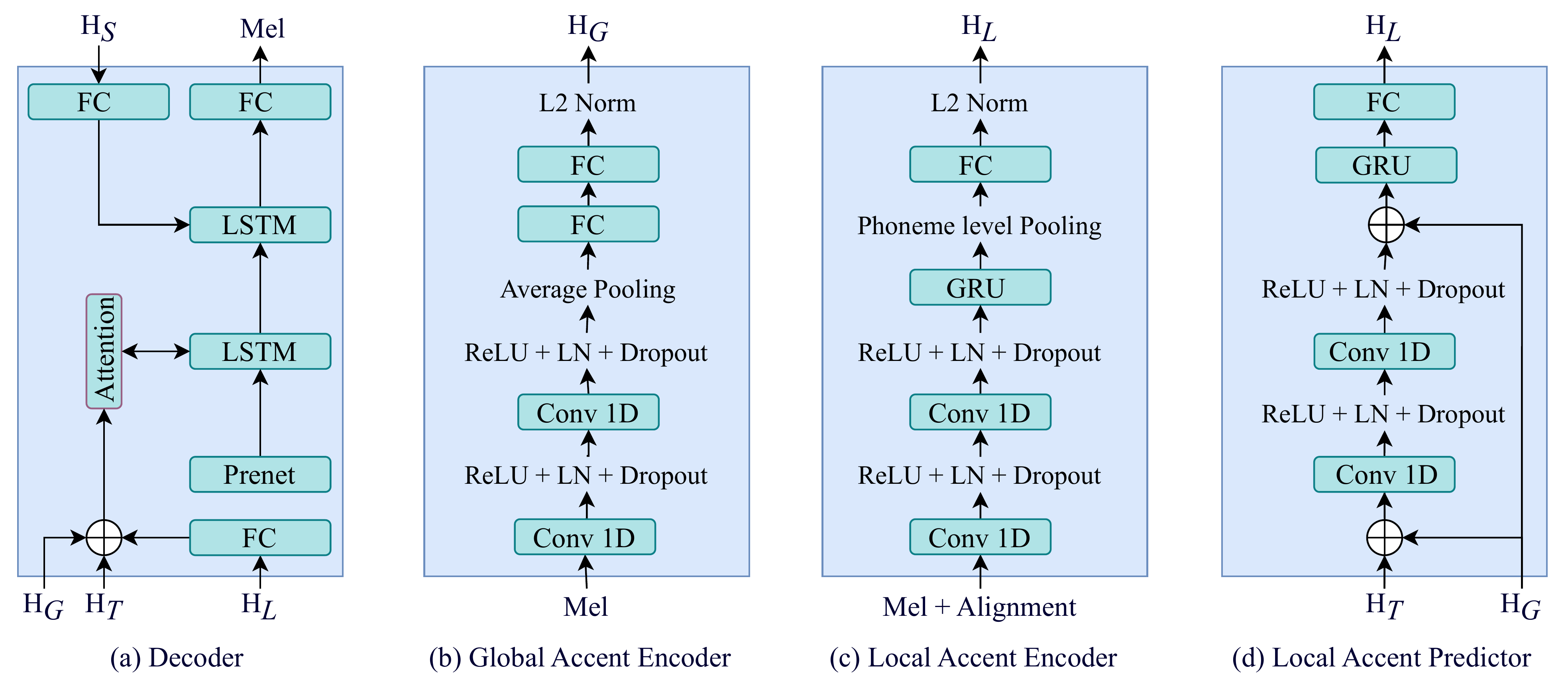}
\caption{The architecture of (a) Decoder, (b) Global Accent Encoder, (c) Local Accent Encoder, (d) Local Accent Predictor. LN denotes layer normalization.}
\label{fig_abcd}
\end{figure*}

\subsection{Acoustic Model (AM)}
We adopt Tacotron 2 \cite{shen2018natural}, an encoder-decoder-based architecture, as our AM. 
The input phoneme sequence is passed to the text encoder, which consists of a phoneme embedding table, three 1-dimensional convolutional layers, and a bi-directional long short-term memory (LSTM) layer. 
The text encoder converts the phoneme sequence into a sequence of hidden text representations, denoted as $H_{T}$. To control accent-specific information in the generated speech, $H_{T}$ is further conditioned on accent representations from the SIGAM, denoted as $H_{G}$, and from the SILAM, denoted as $H_{L}$. 
Both $H_{T}$ and $H_{L}$ have the same length, as they represent phoneme level representations, while $H_{G}$ is a single vector. 
The overall encoding process captures both text representations from phoneme inputs and multi-scale accent representations from accented speech. 

The attention-based decoder is shown in Fig. \ref{fig_abcd}(a). 
$H_{G}$ is first duplicated to match the phoneme level, and $H_{L}$ is projected through a fully connected (FC) layer. These accent representations are then added to $H_{T}$ before being passed to the attention network.
We control the speaker identity in the decoder using the speaker embedding vector $H_{S}$, which serves as an additional input to the decoder. Specifically, $H_{S}$ is transformed through an FC layer with Softsign activation and then concatenated with both the input and output of the second LSTM layer after the attention network at each frame step. This combination mitigates the influence of speaker-specific information on the attention mechanism, enabling it to focus on accent representations to learn accent-specific phoneme durations. The decoder predicts the Mel-spectrogram and stop token label, followed by a postnet that further enhances the Mel-spectrogram prediction, as described in \cite{shen2018natural}. We denote the objective function of the AM as $\mathcal{L}_{Taco2}$. 

\subsection{Speaker-Independent Global Accent Model (SIGAM)} 
The SIGAM aims to capture the overall accent fluctuations of an utterance from the Mel-spectrogram.
The global accent encoder is proposed to generate the utterance level embedding vector $H_{G}$ that serves as a global accent representation.
To make $H_{G}$ be accent-discriminative, the global accent encoder is supervised by an accent classifier.
The architecture of the global accent encoder is shown in Fig. \ref{fig_abcd}(b). The input Mel-spectrogram is passed to two 1-dimensional convolutional layers each with ReLU activation, layer normalization, and dropout. An average pooling operation is applied along the time axis to produce a single vector representing the utterance level speech variation. This vector is then passed through two FC layers to compute $H_{G}$. L2 normalization is subsequently applied to $H_{G}$ to enhance its generalization ability. 
The accent classifier, consisting of an FC layer and a softmax layer, determines the probability distribution of predicted accents. 
The training objective of the accent classifier $\mathcal{L}_{G\_ac}$ is defined as the cross-entropy (CE) loss between the predicted and target accent labels. The CE loss is computed as:
\begin{equation}
\begin{split}
   \mathcal{L}_{CE} = -\sum_{i=1}^{N}log(p(X_i|\hat{X_i})) 
\label{eq:ce}
\end{split}
\end{equation}
where $N$ is the number of categories, and $p(X_i|\widehat{X}_i)$ represents the softmax output, i.e., the probability that the predicted label $\widehat{X}_i$ matches the target label $X_i$. 

To achieve the independent control of accents regardless of speaker characteristics, i.e., speaker-independent accent modeling, the global accent encoder is further adversarially trained with a speaker classifier to disentangle speaker identity from $H_{G}$. 
A gradient reversal layer (GRL) is used between the global accent encoder and the speaker classifier, reversely optimizing the global accent encoder with respect to speaker classification, thereby preventing it from encoding speaker identity. As a result, the SIGAM produces the vector $H_{G}$ that is both speaker-independent and accent-discriminative. 
The speaker classifier consists of an FC layer and a softmax layer to produce the probability distribution of predicted speakers. The loss function of the adversarial speaker classifier $\mathcal{L}_{G\_adv\_sc}$ is defined as the CE loss, as shown in Equation \ref{eq:ce}. 

\subsection{Speaker-Independent Local Accent Model (SILAM)}
Similar to the SIGAM, the SILAM consists of a local accent encoder, an accent classifier, and an adversarial speaker classifier. However, the SILAM is specifically designed to capture fine-grained accent variations, such as pronunciation and prosody on the phoneme level, providing a detailed characterization of segmental accented speech. 
The objective of the SILAM is to produce a sequence of phoneme level embeddings $H_{L}$ that is both speaker-independent and accent-discriminative, serving as local accent representations.

The local accent encoder takes the Mel-spectrogram and force-aligned phoneme boundaries as inputs, as illustrated in Fig. \ref{fig_abcd}(c). It comprises two 1-dimensional convolutional layers each with ReLU activation, layer normalization, and dropout. The output of the second convolutional layer is passed to a gated recurrent unit (GRU) layer to extract frame level acoustic conditions. Phoneme level speech representations are subsequently obtained by applying average pooling over the frame level acoustic conditions for each phoneme, according to phoneme boundaries. To compactly represent phoneme level prosody information \cite{chen2021adaspeech}, the phoneme level speech representations are projected into a low-dimensional space through an FC layer, resulting in $H_{L}$. Finally, L2 normalization is applied to each vector in $H_{L}$ to enhance its predictability. 
 
Since not all phonetic representations exhibit variations across accents, i.e., some phonetic information is shared across accents \cite{gomez2009british}, classifying each component of phoneme level speech representations by accent category may not be optimal. To address this, the accent classifier uses an LSTM layer to capture sequential variations within an utterance. The final state of the LSTM is then passed to an FC layer, followed by a softmax layer to predict the accent probability. 
An adversarial speaker classifier, with the same architecture as that in the SIGAM, is employed in the SILAM to remove speaker information from $H_{L}$. 
By operating on each embedding vector in $H_{L}$, the adversarial speaker classifier ensures that $H_{L}$ becomes speaker-independent. 
Both loss functions of the accent classifier $\mathcal{L}_{L\_ac}$ and adversarial speaker classifier $\mathcal{L}_{L\_adv\_sc}$ are defined as the CE loss, as shown in Equation \ref{eq:ce}. 

\subsection{Local Accent Prediction Model (LAPM)}
The goal of the LAPM is to substitute the SILAM in the inference stage, bypassing the dependency on reference speech. The LAPM consists of a text encoder and a local accent predictor. The text encoder is identical to the one in the AM, producing text representations $H_{T}$, while the local accent predictor generates local accent representations $H_{L}$. As shown in Fig. \ref{fig_abcd}(d), the local accent predictor comprises two 1-dimensional convolutional layers each with ReLU activation, layer normalization, and dropout. These are followed by a GRU layer and an FC layer to predict $H_{L}$.

To enable LAPM predictions across multiple accents, the global accent representation $H_{G}$ from the SIGAM is taken as an additional input to the local accent predictor.
Specifically, $H_{G}$ is added to $H_{T}$ and the input of the GRU layer, respectively.
Overall, the LAPM takes both the phoneme sequence and $H_{G}$ as inputs to predict $H_{L}$, the output of the SILAM. The objective function of the LAPM $\mathcal{L}_{predict}$ is defined as the mean squared error (MSE) loss between the predicted $\hat{H_{L}}$ and target $H_{L}$. 

\subsection{Training Stages}
The proposed TTS framework includes two training stages. 
In the first stage, the AM, SIGAM, and SILAM are jointly trained using the total objective function defined as:
\begin{equation}
\begin{aligned}
    \mathcal{L}_{train\_TTS} = \alpha\mathcal{L}_{Taco2} + \beta\mathcal{L}_{G\_ac} &+ \gamma\mathcal{L}_{G\_adv\_sc} \\ + \delta\mathcal{L}_{L\_ac} &+ \epsilon\mathcal{L}_{L\_adv\_sc}
\end{aligned}
\label{equ_tts}
\end{equation}
where $\alpha$, $\beta$, $\gamma$, $\delta$, and $\epsilon$ are parameters to balance the weights of different losses.
In the second stage, only the LAPM is trained using the objective function $\mathcal{L}_{predict}$. Prior to training the LAPM, we extract $H_{G}$ and $H_{L}$ from the training data using the SIGAM and SILAM, respectively, which are trained in the first stage. The text encoder in the LAPM shares the same weights as the one in the AM trained in the first stage and is frozen during the LAPM training. 

\subsection{Inference Stage}
Our framework generates accented speech directly from the phoneme sequence using the LAPM, as shown in Fig. \ref{fig_sys}. 
By disentangling speaker information within the SIGAM, all utterance level $H_{G}$ vectors from different utterances converge closely within the same accent. Therefore, we use a single embedding vector $H_{Avg\_G}$ to represent each accent category during inference. Specifically, $H_{Avg\_G}$ is computed as the average of all $H_{G}$ vectors extracted from the training utterances of the corresponding accent. 

\section{Experimental Setup}
\label{sec:exp_setup}

\subsection{Database}
We use the multi-speaker accented English speech corpus, L2-ARCTIC \cite{zhao2018l2}, for all experiments. This corpus contains recordings from 24 foreign-accented speakers across six accents: Arabic (AR), Mandarin (ZH), Hindi (HI), Korean (KO), Spanish (ES), and Vietnamese (VI), with four speakers per accent. The dataset is divided into 23,075 training utterances, 1,200 validation utterances (50 per speaker), and 2,400 test utterances (100 per speaker). The text transcriptions in the L2-ARCTIC corpus are parallel across different speakers, except for a few utterances. Phoneme sequences and force-aligned phoneme boundaries are provided by the corpus. We trim the silence at the beginning and end of each utterance. All speech signals are downsampled to 16 kHz, and the 80-dimensional Mel-spectrogram is extracted with a 50 ms frame length and a 12.5 ms frame shift.

Our experiments focus on generating multi-speaker multi-accent speech, which involves two scenarios: generating the voices of multiple speakers with their own accents, (\emph{multi-speaker inherent-accent}) and with other different accents (\emph{multi-speaker cross-accent}). 
All target speakers are from the L2-ARCTIC corpus, where each speaker has only one accent. For \emph{multi-speaker cross-accent} speech synthesis, 
we randomly select a male speaker \textit{BWC} and a female speaker \textit{LXC} from the ZH accent as target speakers, while the remaining five accents are regarded as target accents to be generated.

\subsection{Implementations}
Our AM architecture follows Tacotron 2 \cite{shen2018natural}. Each TTS system is trained for 600k steps, and the LAPM is trained for 200k steps.
All systems are optimized using the Adam optimizer \cite{kingma2014adam} with a batch size of 32. The initial learning rate is set to 1e-3 and gradually decays to 1e-5. The parameters in Equation \ref{equ_tts} are set to $\alpha$ = 1, $\beta$ = 1, $\gamma$ = 0.02, $\delta$ = 1, and $\epsilon$ = 0.02. The 256-dimensional speaker embedding, extracted from a pre-trained speaker encoder\footnote{\url{https://github.com/resemble-ai/Resemblyzer}}, is combined with the decoder in the same way for all compared TTS systems. 
To ensure fair comparisons, we use the same neural vocoder, Parallel WaveGAN \cite{yamamoto2020parallel}, to generate speech waveforms from the predicted Mel-spectrogram for all TTS systems. The Parallel WaveGAN is pre-trained on the CSTR\_VCTK \cite{veaux2017cstr} speech corpus. 
The following TTS systems are implemented for experiments: 

\begin{itemize}
\item \textbf{AM-GST:} 
A multi-speaker TTS system that conditions the AM on the GST model \cite{wang2018style}. We set the number of token layers to six, corresponding to the number of accent categories during training. 

\item \textbf{AM-VAE:}
A multi-speaker TTS system that conditions the AM on the VAE model \cite{zhang2019learning}. 

\item \textbf{AM-SIGAM:}
A multi-speaker TTS system that conditions the AM on the SIGAM. 

\item \textbf{AM-SIMSAM:}
A multi-speaker TTS system that conditions the AM on the speaker-independent multi-scale accent model (SIMSAM), which includes the SIGAM, SILAM, and LAPM, as shown in Fig. \ref{fig_sys}. 

\item \textbf{AM-MSAM:}
A multi-speaker TTS system that conditions the AM on the multi-scale accent model (MSAM), consisting of the global accent model (GAM) and local accent model (LAM), 
without the speaker disentanglement. 
Predicting phoneme level $H_{L}$ directly from the phoneme sequence is challenging due to the entanglement of speaker information in $H_{L}$. Therefore, the LAPM is excluded from this system, and reference speech is required during inference. 
\end{itemize}

Note that for both GST and VAE models, the average of all utterance level embeddings extracted from the training data of each accent is used to represent the corresponding accent category during inference, similar to $H_{Avg\_G}$ for the SIGAM. 

\subsection{Evaluation Metrics}
We utilize both objective and subjective evaluation metrics to assess the generated accented speech. 

\subsubsection{Objective evaluations} 
When the ground truth of the generated speech is available, we objectively evaluate the system performance. Before the evaluations, dynamic time warping (DTW) \cite{muller2007dynamic} is used to align the generated speech and ground truth to the same length. 
We utilize Mel-cepstral distortion (MCD) \cite{kubichek1993mel} to evaluate speech quality. MCD measures the distance between the Mel-cepstrum extracted from the generated speech and ground truth. A lower MCD value indicates better speech quality. 
Accent similarity is evaluated based on two important elements of accents: pitch and duration. 
To assess pitch variations, we calculate root mean squared error (RMSE) \cite{wang2018autoregressive} and Pearson's correlation coefficient \cite{cohen2009pearson} of the fundamental frequency ($F0$), where the entire $F0$ sequence is used for the evaluations. We extract the $F0$ from speech waveforms using pyworld\footnote{\url{https://pypi.org/project/pyworld/}}. 
A lower $F0$ RMSE and a higher $F0$ correlation indicate better pitch prediction.
Duration is assessed using frame disturbance (FD) \cite{liu2021expressive} on the aligned path of the DTW results. A lower FD value suggests more accurate duration reconstruction. 

For speaker similarity evaluation, we calculate the cosine similarity \cite{kim2020emotional} between utterance level speaker embeddings (SECS) extracted from two speech samples. A value closer to 1 indicates higher speaker similarity. Note that this metric can be used even when the ground truth is unavailable. 
We also calculate the cosine similarity between utterance level accent embeddings (AECS) extracted from the ground truth accented speech to evaluate accent similarity in Section \ref{sec:ablation_study_speaker_disentanglement}. 

\subsubsection{Subjective evaluations}
We conduct subjective evaluations through listening tests. 20 participants from the United States are recruited on Amazon Mechanical Turk\footnote{\url{https://www.mturk.com}} for each listening test\footnote{All speech samples are available at: \url{https://xuehao-marker.github.io/MSMA-TTS/}}, and they are compensated upon completing the tasks. 
In the listening experiments for each accent, six groups of utterances are randomly selected from the test set for evaluations. 
We use the mean opinion score (MOS) \cite{streijl2016mean} test to evaluate speech quality in terms of naturalness (NMOS).
In the NMOS test, participants rate the provided speech samples based on speech naturalness. The optional score ranges from 1 to 5 with intervals of 0.5, where 1 = bad, 2 = poor, 3 = fair, 4 = good, and 5 = excellent. 
The MOS test is also used to evaluate accent similarity (AMOS) and speaker similarity (SMOS).
In the AMOS and SMOS tests, participants first listen to the reference speech and then rate the provided speech samples only according to accent or speaker similarity compared to the reference speech. The scoring range is the same as that in the NMOS test. 

Additionally, we conduct XAB preference tests to further evaluate accent and speaker similarity.
In these tests, participants first listen to the reference speech X and then select a speech sample, from A and B, that is more similar to the reference speech only according to accent or speaker similarity. In all accent and speaker similarity tests, participants are instructed to ignore the speech content and quality. 

\section{Experimental Results}
\label{sec:exp_results}

\subsection{Comparisons with Baseline Systems}
We compare our proposed system, AM-SIMSAM, with two baseline systems, AM-GST and AM-VAE, in both \emph{multi-speaker inherent-accent} and \emph{multi-speaker cross-accent} speech synthesis scenarios to comprehensively evaluate performance. 
\subsubsection{Multi-speaker inherent-accent speech synthesis}

\begin{table*}
\caption{Results of MCD for speech quality, $F0$ RMSE and $F0$ correlation for pitch, and FD for duration.}
\centering
\label{tab:mcd_f0_dur_inherent}
\renewcommand\arraystretch{1}{
\scalebox{0.79}{
\begin{tabular}{p{1cm}<{\centering}|p{0.9cm}<{\centering}p{0.9cm}<{\centering}p{0.9cm}<{\centering}p{0.95cm}<{\centering}|p{0.9cm}<{\centering}p{0.9cm}<{\centering}p{0.9cm}<{\centering}p{0.95cm}<{\centering}|p{0.9cm}<{\centering}p{0.9cm}<{\centering}p{0.9cm}<{\centering}p{0.95cm}<{\centering}|p{0.9cm}<{\centering}p{0.9cm}<{\centering}p{0.9cm}<{\centering}p{0.95cm}<{\centering}} 

\hline
\specialrule{0em}{0pt}{2pt}
  & \multicolumn{4}{c|}{MCD (dB)} & \multicolumn{4}{c|}{$F0$ RMSE (Hz)} & \multicolumn{4}{c|}{$F0$ correlation}  & \multicolumn{4}{c}{FD (Frame)} \\
\specialrule{0em}{2pt}{0pt}
\cline{2-17}
\specialrule{0em}{0pt}{2pt}
         & AM-GST  & AM-VAE  &AM-SIGAM  & AM-SIMSAM & AM-GST  & AM-VAE  &AM-SIGAM & AM-SIMSAM & AM-GST  & AM-VAE &AM-SIGAM & AM-SIMSAM & AM-GST  & AM-VAE  &AM-SIGAM  & AM-SIMSAM \\ 
\specialrule{0em}{2pt}{0pt}
         \hline
\specialrule{0em}{0pt}{2pt}     
\multicolumn{17}{c}{\emph{Multi-speaker inherent-accent}} \\ 
\specialrule{0em}{0pt}{2pt}
\hline
\specialrule{0em}{0pt}{2pt}
AR     & 8.33 & 8.33 &8.24 &\textbf{8.04}        & 70.27 & 68.09   & 68.77 & \textbf{66.86}         & 0.665 & 0.681  &0.681  & \textbf{0.697}                   & 25.29 & 21.96  &21.37&  \textbf{17.41}    \\ 
ZH     & 8.12 & 8.10 &7.95 &\textbf{7.83}      & 57.95 & 56.34 &56.37 & \textbf{54.88}              & 0.705 & 0.720 &0.721 & \textbf{0.736}                    & 27.83 & 23.22 &21.54 &\textbf{19.50}       \\ 
HI   & 8.20 & 8.11 &8.11 &\textbf{7.92}          &  74.86 & 74.17  & 75.02& \textbf{73.45}          & 0.654 & 0.669  & 0.662& \textbf{0.679}                     & 20.68 & 17.15  & 16.68& \textbf{14.03}    \\  
KO        & 8.17 & 8.26 &8.15 &\textbf{8.02}      & 68.39 & 66.38  & 67.00& \textbf{65.11}           & 0.697 & 0.704  &0.699& \textbf{0.718}                    & 21.33 & 17.08 &15.73 &\textbf{15.40}     \\  
ES    & 7.60 & 7.85 &7.76 &\textbf{7.54}        & 62.02 & 62.80 &63.30& \textbf{61.11}              & 0.680 & 0.681  &0.684& \textbf{0.699}                & 31.08 & 28.23  &27.71& \textbf{23.50}       \\ 
VI     & 8.00 & 7.91  &7.69&\textbf{7.34}      & 66.94 & 66.00  &64.84&  \textbf{62.29}          & 0.693 & 0.705  & 0.714& \textbf{0.737}             & 27.47 & 25.19 &24.68&\textbf{20.38}                \\ 
\specialrule{0em}{2pt}{0pt}
\hline
\specialrule{0em}{0pt}{2pt}
AVG       & 8.07 & 8.10 &7.98&\textbf{7.78}       & 66.74 & 65.63 &65.88 & \textbf{63.95}            & 0.682 & 0.693  &0.692 & \textbf{0.711}                 & 25.61 & 22.14  &21.29 &\textbf{18.37}     \\ 
\specialrule{0em}{2pt}{0pt}
\hline
\end{tabular}}}
\end{table*}

\begin{table}
\caption{Results of SECS for speaker similarity.}
\centering
\label{tab:secs_inherent_cross}
\renewcommand\arraystretch{1}{
\scalebox{0.91}{
\begin{tabular}{p{1cm}<{\centering}|p{0.9cm}<{\centering}p{0.9cm}<{\centering}p{0.95cm}<{\centering}|p{0.9cm}<{\centering}p{0.9cm}<{\centering}p{0.95cm}<{\centering}} 
\hline
\specialrule{0em}{2pt}{0pt}
&\multicolumn{3}{c|}{\emph{Multi-speaker inherent-accent}} & \multicolumn{3}{c}{\emph{Multi-speaker cross-accent}} \\ 
\specialrule{0em}{2pt}{0pt}
\cline{2-7}
\specialrule{0em}{2pt}{0pt}
& AM-GST & AM-VAE & AM-SIMSAM & AM-GST & AM-VAE & AM-SIMSAM \\
\specialrule{0em}{2pt}{0pt}
\hline
\specialrule{0em}{2pt}{0pt}
AR  & 0.888 & 0.89  & \textbf{0.899}   & 0.866 & \textbf{0.907} & 0.857   \\ 
ZH  & 0.893 & 0.907 & \textbf{0.913}   & -     & -              & -       \\ 
HI  & 0.868 & 0.885 & \textbf{0.895}   & 0.868 & \textbf{0.904} & 0.842   \\
KO  & 0.906 & 0.902 & \textbf{0.907}   & 0.865 & \textbf{0.905} & 0.849   \\
ES  & 0.908 & 0.906 & \textbf{0.911}   & \textbf{0.908} & 0.906 & 0.879   \\
VI  & 0.888 & 0.901 & \textbf{0.904}   & 0.873 & \textbf{0.904} & 0.858   \\
\specialrule{0em}{2pt}{0pt}
\hline
\specialrule{0em}{2pt}{0pt}
AVG & 0.892 & 0.899 & \textbf{0.905}   & 0.876 & \textbf{0.905} & 0.857   \\ 
\specialrule{0em}{2pt}{0pt}
\hline
\end{tabular}}}
\end{table}

\begin{table*}
\caption{Results of NMOS test for speech quality, AMOS test for accent similarity, and SMOS test for speaker similarity. All presented scores are with 95\% confidence intervals.}
\centering
\label{tab:nmos_amos_smos_inherent_cross}
\renewcommand\arraystretch{1}{
\scalebox{0.88}{
\begin{tabular}{p{1cm}<{\centering}|p{1.7cm}<{\centering}p{1.7cm}<{\centering}p{1.7cm}<{\centering}|p{1.7cm}<{\centering}p{1.7cm}<{\centering}p{1.7cm}<{\centering}|p{1.7cm}<{\centering}p{1.7cm}<{\centering}p{1.7cm}<{\centering}} 

\hline
\specialrule{0em}{0pt}{2pt}
  & \multicolumn{3}{c|}{NMOS} & \multicolumn{3}{c|}{AMOS} & \multicolumn{3}{c}{SMOS} \\ 
\specialrule{0em}{2pt}{0pt}
\cline{2-10}
\specialrule{0em}{0pt}{2pt}
         & AM-GST  & AM-VAE    & AM-SIMSAM & AM-GST  & AM-VAE   & AM-SIMSAM & AM-GST  & AM-VAE  & AM-SIMSAM  \\ 
\specialrule{0em}{2pt}{0pt}
         \hline
         \specialrule{0em}{2pt}{0pt}
\multicolumn{10}{c}{\emph{Multi-speaker inherent-accent}} \\ 
\specialrule{0em}{2pt}{0pt} \hline
\specialrule{0em}{0pt}{2pt}      
AR     & 3.81 $\pm$ 0.17 & 3.73 $\pm$ 0.16  &\textbf{3.87 $\pm$ 0.14}        & 3.70 $\pm$ 0.18 & 3.79 $\pm$ 0.18 & \textbf{3.89 $\pm$ 0.16}         & 4.00 $\pm$ 0.17 & 3.94 $\pm$ 0.17 & \textbf{4.04 $\pm$ 0.17}       \\ 
ZH     & 3.58 $\pm$ 0.17 & 3.49 $\pm$ 0.17  &\textbf{3.73 $\pm$ 0.14}        & 3.67 $\pm$ 0.19 & 3.65 $\pm$ 0.19 & \textbf{3.73 $\pm$ 0.17}         & \textbf{3.78 $\pm$ 0.17} & 3.71 $\pm$ 0.17 & 3.77 $\pm$ 0.18       \\ 
HI     & 3.75 $\pm$ 0.15 & 3.77 $\pm$ 0.16  &\textbf{3.83 $\pm$ 0.15}        & 3.88 $\pm$ 0.17 & 3.93 $\pm$ 0.17 & \textbf{4.00 $\pm$ 0.16}         & 3.63 $\pm$ 0.19 & 3.69 $\pm$ 0.17 & \textbf{3.72 $\pm$ 0.18}       \\ 
KO     & 3.75 $\pm$ 0.16 & 3.79 $\pm$ 0.15  &\textbf{3.89 $\pm$ 0.16}        & 3.74 $\pm$ 0.18 & 3.75 $\pm$ 0.18 & \textbf{3.79 $\pm$ 0.17}         & 3.75 $\pm$ 0.18 & 3.79 $\pm$ 0.18 & \textbf{3.81 $\pm$ 0.17}       \\  
ES     & 3.67 $\pm$ 0.16 & 3.80 $\pm$ 0.16  &\textbf{3.85 $\pm$ 0.15}        & 3.74 $\pm$ 0.18 & 3.76 $\pm$ 0.20 & \textbf{3.85 $\pm$ 0.18}         & \textbf{3.84 $\pm$ 0.15} & 3.82 $\pm$ 0.15 & 3.83 $\pm$ 0.16       \\
VI     & 3.58 $\pm$ 0.18 & 3.57 $\pm$ 0.19  &\textbf{3.72 $\pm$ 0.19}        & 3.65 $\pm$ 0.19 & 3.70 $\pm$ 0.19 & \textbf{3.77 $\pm$ 0.19}         & \textbf{3.73 $\pm$ 0.17} & 3.73 $\pm$ 0.17 & 3.72 $\pm$ 0.17       \\ 
\specialrule{0em}{2pt}{0pt}
\hline
\specialrule{0em}{2pt}{0pt}
AVG    & 3.69 $\pm$ 0.17 & 3.69 $\pm$ 0.17  &\textbf{3.82 $\pm$ 0.16}        & 3.73 $\pm$ 0.18 & 3.76 $\pm$ 0.19 & \textbf{3.84 $\pm$ 0.17}         & 3.79 $\pm$ 0.17 & 3.78 $\pm$ 0.17 & \textbf{3.82 $\pm$ 0.17}       \\
\specialrule{0em}{2pt}{0pt}
\hline
\specialrule{0em}{2pt}{0pt}
\multicolumn{10}{c}{\emph{Multi-speaker cross-accent}} \\ 
\specialrule{0em}{2pt}{0pt} \hline
\specialrule{0em}{0pt}{2pt}      
AR     & 3.66 $\pm$ 0.16 & 3.32 $\pm$ 0.16  &\textbf{3.80 $\pm$ 0.16}        &  2.69 $\pm$ 0.21 & 2.29 $\pm$ 0.19 & \textbf{3.25 $\pm$ 0.21}       & 3.27 $\pm$ 0.20 & \textbf{3.57 $\pm$ 0.17} & 3.35 $\pm$ 0.19       \\  
HI     & 3.71 $\pm$ 0.15 & 3.10 $\pm$ 0.16  &\textbf{3.78 $\pm$ 0.14}        &  2.93 $\pm$ 0.22 & 2.41 $\pm$ 0.23 & \textbf{3.35 $\pm$ 0.23}       & 3.40 $\pm$ 0.18 & \textbf{3.83 $\pm$ 0.15} & 3.30 $\pm$ 0.20       \\ 
KO     & 3.62 $\pm$ 0.14 & 3.38 $\pm$ 0.16  &\textbf{3.76 $\pm$ 0.15}        &  2.89 $\pm$ 0.21 & 2.35 $\pm$ 0.21 & \textbf{3.09 $\pm$ 0.22}       & 3.20 $\pm$ 0.23 & \textbf{3.70 $\pm$ 0.18} & 3.35 $\pm$ 0.20       \\  
ES     & 3.54 $\pm$ 0.16 & 3.25 $\pm$ 0.18  &\textbf{3.75 $\pm$ 0.16}        &  2.83 $\pm$ 0.21 & 2.73 $\pm$ 0.23 & \textbf{3.32 $\pm$ 0.20}       & 3.66 $\pm$ 0.18 & \textbf{3.73 $\pm$ 0.17} & 3.48 $\pm$ 0.20       \\
VI     & 3.43 $\pm$ 0.17 & 3.16 $\pm$ 0.17  &\textbf{3.48 $\pm$ 0.17}        &  3.03 $\pm$ 0.21 & 2.88 $\pm$ 0.23 & \textbf{3.30 $\pm$ 0.21}       & 3.30 $\pm$ 0.21 & \textbf{3.63 $\pm$ 0.20} & 3.17 $\pm$ 0.21       \\
\specialrule{0em}{2pt}{0pt}
\hline
\specialrule{0em}{2pt}{0pt}
AVG   & 3.59 $\pm$ 0.16 & 3.24 $\pm$ 0.17 & \textbf{3.71 $\pm$ 0.16}         &  2.87 $\pm$ 0.21 & 2.53 $\pm$ 0.22 & \textbf{3.26 $\pm$ 0.21}       & 3.37 $\pm$ 0.20 & \textbf{3.69 $\pm$ 0.17} & 3.33 $\pm$ 0.20       \\
\specialrule{0em}{2pt}{0pt}
\hline
\end{tabular}}}
\end{table*}

Objective evaluations are first conducted in this scenario using the available ground truth. The results of MCD, $F0$ RMSE, $F0$ correlation, and FD are shown in Table \ref{tab:mcd_f0_dur_inherent}, and SECS results are presented in Table \ref{tab:secs_inherent_cross}.
We make the following observations: 1) AM-SIMSAM outperforms both AM-GST and AM-VAE in terms of MCD, suggesting that AM-SIMSAM generates speech with improved quality. 
2) AM-SIMSAM achieves the lowest $F0$ RMSE and the highest $F0$ correlation among the three systems. This demonstrates that the SIMSAM extracts detailed prosody information from the Mel-spectrogram, 
resulting in more accurate pitch prediction in the generated accented speech.
3) In the duration evaluation on FD, AM-SIMSAM exhibits significantly better performance than AM-VAE, followed by AM-GST, indicating that the SIMSAM enhances duration reconstruction of accented speech. 
The results for pitch and duration demonstrate that the SIMSAM effectively enhances prosodic rendering of accents. 
4) AM-SIMSAM achieves the highest scores among the three systems in terms of SECS, showing that it preserves target speaker identity well in the generated speech.
These observations are consistent across all accents, although the extent of improvement varies, likely due to differences in accent variations. 

Next, we conduct subjective evaluations, and the results of the NMOS, AMOS, and SMOS tests are presented in Table \ref{tab:nmos_amos_smos_inherent_cross}.  
We are pleased to observe that AM-SIMSAM achieves the highest scores among the three systems across all accents in both NMOS and AMOS tests, demonstrating the effectiveness of the SIMSAM on improving both speech quality and accent rendering. 
These results indicate that while the GST and VAE models perform well for style modeling in expressive TTS, they are less capable of capturing fine-grained accent variations compared to the SIMSAM.
In the SMOS test, all systems exhibit comparable performance, with AM-SIMSAM achieving slightly higher scores than the others, suggesting its capability to generate multi-speaker speech with high speaker similarity. 
Overall, both objective and subjective evaluations demonstrate that AM-SIMSAM has the best performance among the three systems for \emph{multi-speaker inherent-accent} speech synthesis. 

\subsubsection{Multi-speaker cross-accent speech synthesis} 
In this scenario, where the ground truth is unavailable, subjective metrics are used as the primary evaluation methods. 
The results of the NMOS, AMOS, and SMOS tests for this scenario are shown in the lower part of Table \ref{tab:nmos_amos_smos_inherent_cross}. 
It is observed that the scores of these three tests are lower compared to those in the \emph{multi-speaker inherent-accent} scenario, as the accents in the generated speech are unseen to the target speakers in this scenario. 
In the NMOS test, AM-SIMSAM consistently achieves the highest scores among the three systems across all accents, confirming that the SIMSAM contributes to higher quality and more natural speech. 
A similar trend is observed in the AMOS test, where system performance in terms of accent similarity is ranked in descending order: AM-SIMSAM, AM-GST, and AM-VAE.
Notably, AM-SIMSAM achieves significantly higher AMOS scores than the other two systems, with relative improvements being substantially greater compared to those observed in the \emph{multi-speaker inherent-accent} scenario. 
These results strongly validate the capability of the SIMSAM to capture complex accent characteristics within a TTS system, enabling accurate and effective accent rendering, particularly for generating \emph{multi-speaker cross-accent} speech.
In contrast, AM-VAE achieves the lowest AMOS scores, 
suggesting the limitations of the latent representation learned by the VAE model in representing accents, leading to weak accent rendering in the generated speech.

Regarding speaker similarity, the SECS results in Table \ref{tab:secs_inherent_cross} indicate that AM-VAE achieves the highest speaker similarity among the three systems. Similarly, 
the SMOS test results in Table \ref{tab:nmos_amos_smos_inherent_cross} show that AM-VAE significantly outperforms both AM-GST and AM-SIMSAM on target speaker similarity. 
However, despite this advantage, AM-VAE is not suitable for this scenario due to its lowest NMOS scores and significant limitations in accent rendering. 
Overall, AM-SIMSAM is the preferred system for \emph{multi-speaker cross-accent} speech synthesis, demonstrating superior performance in terms of speech quality and accent rendering, despite some compromises on target speaker similarity. 

In conclusion, the results in both scenarios demonstrate that AM-SIMSAM is a versatile and effective solution for multi-speaker multi-accent speech synthesis, outperforming two baseline systems on speech quality and accent similarity. 



\subsection{Ablation Studies}
We conduct ablation studies to evaluate the effectiveness of different components of our proposed system: the SIGAM, SILAM with LAPM, and speaker disentanglement.

\begin{table*}
\caption{Results of accent XAB preference test for accent similarity. NP denotes No Preference.}
\centering
\label{tab:xab_accent_inherent_cross}
\renewcommand\arraystretch{1}{
\scalebox{0.88}{
\begin{tabular}{p{1cm}<{\centering}|p{1.7cm}<{\centering}p{1.7cm}<{\centering}p{1.7cm}<{\centering}|p{1.7cm}<{\centering}p{1.7cm}<{\centering}p{1.7cm}<{\centering}|p{1.7cm}<{\centering}p{1.7cm}<{\centering}p{1.7cm}<{\centering}} 
\hline
\specialrule{0em}{0pt}{2pt}
 & \multicolumn{9}{c}{Accent XAB Preference (\%)}  \\ 
\specialrule{0em}{2pt}{0pt}
\cline{2-10}
\specialrule{0em}{2pt}{0pt}
         & AM-GST  & NP    & AM-SIGAM & AM-VAE  & NP   & AM-SIGAM & AM-SIGAM  & NP  & AM-SIMSAM  \\ 
\specialrule{0em}{2pt}{0pt}
         \hline
         \specialrule{0em}{2pt}{0pt}
\multicolumn{10}{c}{\emph{Multi-speaker inherent-accent}} \\ 
\specialrule{0em}{2pt}{0pt} \hline
\specialrule{0em}{0pt}{2pt}      
AR     & 34.17 & 19.16 & \textbf{46.67}        & 31.67 & 12.50 & \textbf{55.83}         & 27.50 & 15.83 & \textbf{56.67}                   \\ 
ZH     & 26.67 & 15.83 & \textbf{57.50}        & 33.33 & 20.00 & \textbf{46.67}         & 31.67 & 21.66 & \textbf{46.67}                   \\ 
HI     & 26.67 & 17.50 & \textbf{55.83}        & 35.00 & 23.33 & \textbf{41.67}         & 27.50 & 15.00 & \textbf{57.50}                   \\  
KO     & 38.33 & 12.50 & \textbf{49.17}        & 33.33 & 26.67 & \textbf{40.00}         & 33.33 & 18.34 & \textbf{48.33}                   \\  
ES     & 30.83 & 22.50 & \textbf{46.67}        & 35.00 & 24.17 & \textbf{40.83}         & 29.17 & 16.66 & \textbf{54.17}                   \\ 
VI     & 30.83 & 24.17 & \textbf{45.00}        & 32.50 & 27.50 & \textbf{40.00}         & 28.33 & 18.34 & \textbf{53.33}                   \\ 
\specialrule{0em}{2pt}{0pt}
\hline
\specialrule{0em}{2pt}{0pt}
AVG    & 31.25 & 18.61 & \textbf{50.14}        & 33.47 & 22.36 & \textbf{44.17}         & 29.58 & 17.64 & \textbf{52.78}                  \\
\specialrule{0em}{2pt}{0pt}
\hline
         \specialrule{0em}{2pt}{0pt}
\multicolumn{10}{c}{\emph{Multi-speaker cross-accent}} \\ 
\specialrule{0em}{2pt}{0pt} \hline
\specialrule{0em}{0pt}{2pt}      
AR     & 37.50 & 14.17 &\textbf{48.33}        & 34.17 & 13.33 & \textbf{52.50}         & 24.17 & 15.00 & \textbf{60.83}                   \\  
HI     & 33.33 & 19.17 &\textbf{47.50}        & 28.33 & 15.00 & \textbf{56.67}         & 23.33 & 20.00 & \textbf{56.67}                   \\ 
KO     & 38.33 & 20.84 &\textbf{40.83}        & 28.33 & 22.50 & \textbf{49.17}         & 29.17 & 15.00 & \textbf{55.83}                   \\ 
ES     & 36.67 & 18.33 &\textbf{45.00}        & 38.33 & 20.00 & \textbf{41.67}         & 27.50 & 17.50 & \textbf{55.00}                   \\ 
VI     & 36.67 & 21.66 &\textbf{41.67}        & 35.83 & 19.17 & \textbf{45.00}         & 23.33 & 16.67 & \textbf{60.00}                   \\ 
\specialrule{0em}{2pt}{0pt}
\hline
\specialrule{0em}{2pt}{0pt}
AVG    & 36.50 & 18.83 & \textbf{44.67}       & 33.00 & 18.00 & \textbf{49.00}         & 25.50 & 16.83 & \textbf{57.67}                   \\
\specialrule{0em}{2pt}{0pt}
\hline
\end{tabular}}}
\end{table*}

\subsubsection{Effectiveness of the SIGAM}
The impact of the SIGAM on accent rendering is evaluated by comparing AM-SIGAM with AM-GST and AM-VAE, as all address accents at the global scale.
The accent XAB preference test results are presented in Table \ref{tab:xab_accent_inherent_cross}. 
We are glad to see that AM-SIGAM consistently achieves higher accent preference scores than both AM-GST and AM-VAE in both scenarios. This suggests the effectiveness of the SIGAM on enhancing accent rendering.
Objective evaluation results in Table \ref{tab:mcd_f0_dur_inherent} show that AM-SIGAM achieves lower FD than AM-VAE, followed by AM-GST, for duration.
However, we note that on $F0$ RMSE and $F0$ correlation for pitch, AM-SIGAM performs better than AM-GST but slightly worse than AM-VAE. We suspect that the higher accent preference scores of AM-SIGAM in subjective evaluations may be influenced by multiple factors, such as pronunciation and speech duration. 
Additionally, the lower MCD achieved by AM-SIGAM compared to both AM-GST and AM-VAE indicates better speech quality. 


\begin{figure*}
\centering
\begin{minipage}{.32\textwidth}
    \centerline{\includegraphics[width=5.7cm, height=3.5cm]{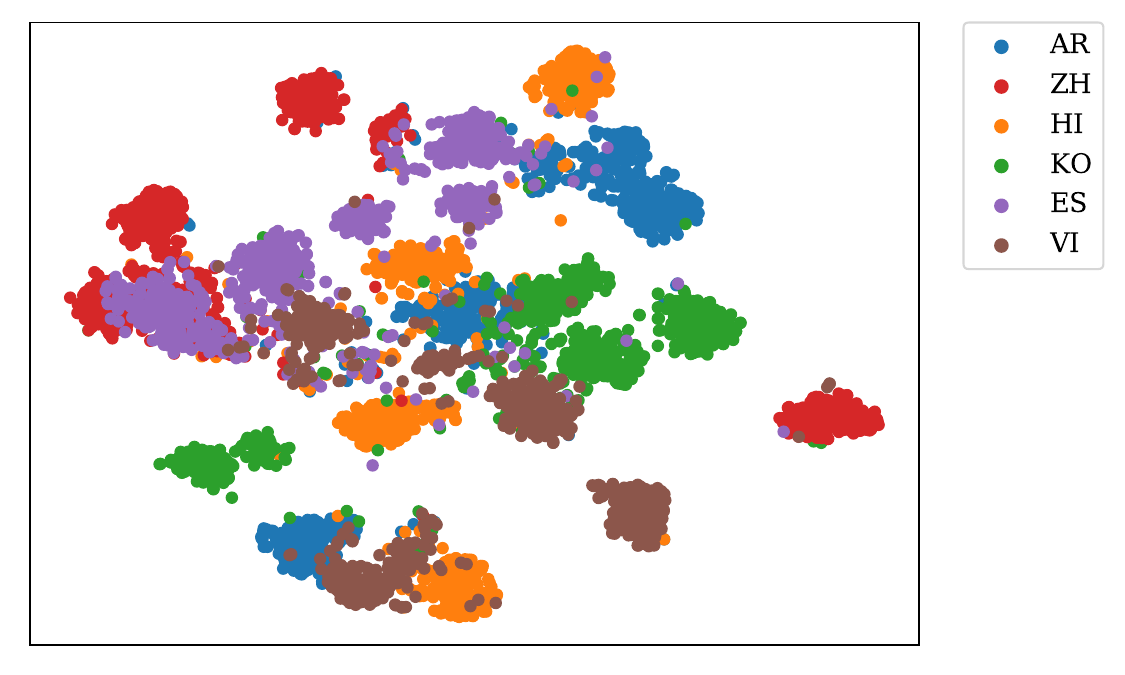}}
    \centerline{(a) GST}
    \label{fig:a}
\end{minipage}
\begin{minipage}{.32\textwidth}
    \centerline{\includegraphics[width=5.7cm, height=3.5cm]{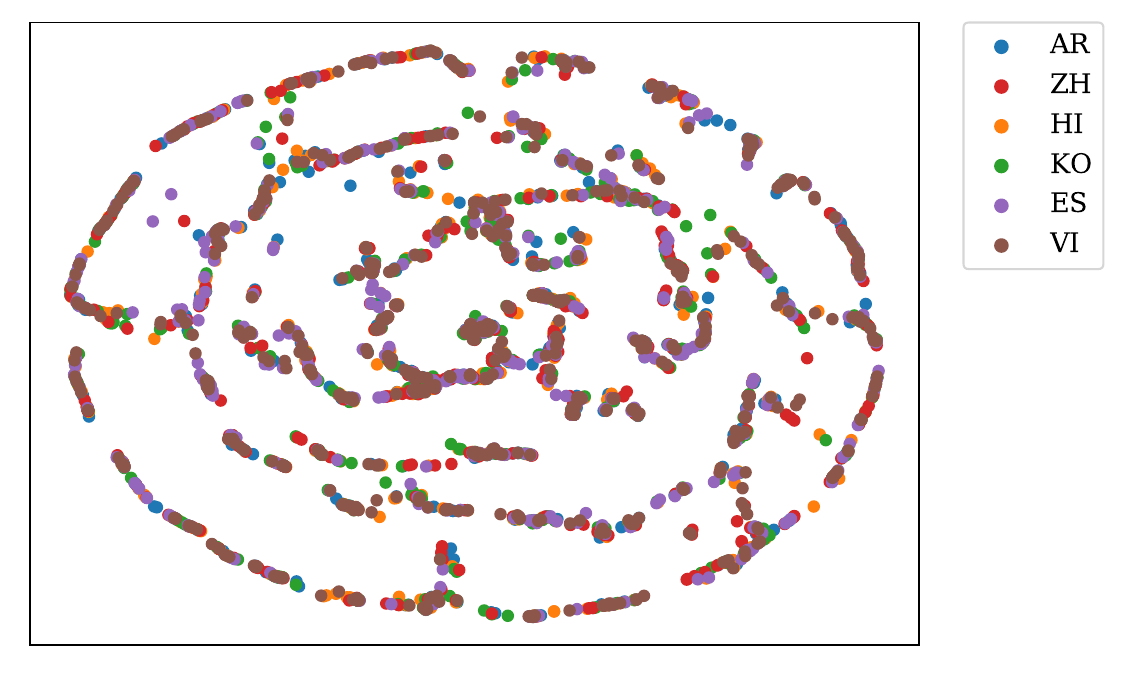}}
    \centerline{(b) VAE}
    \label{fig:b}
\end{minipage}
\begin{minipage}{.32\textwidth}
    \centerline{\includegraphics[width=5.7cm, height=3.5cm]{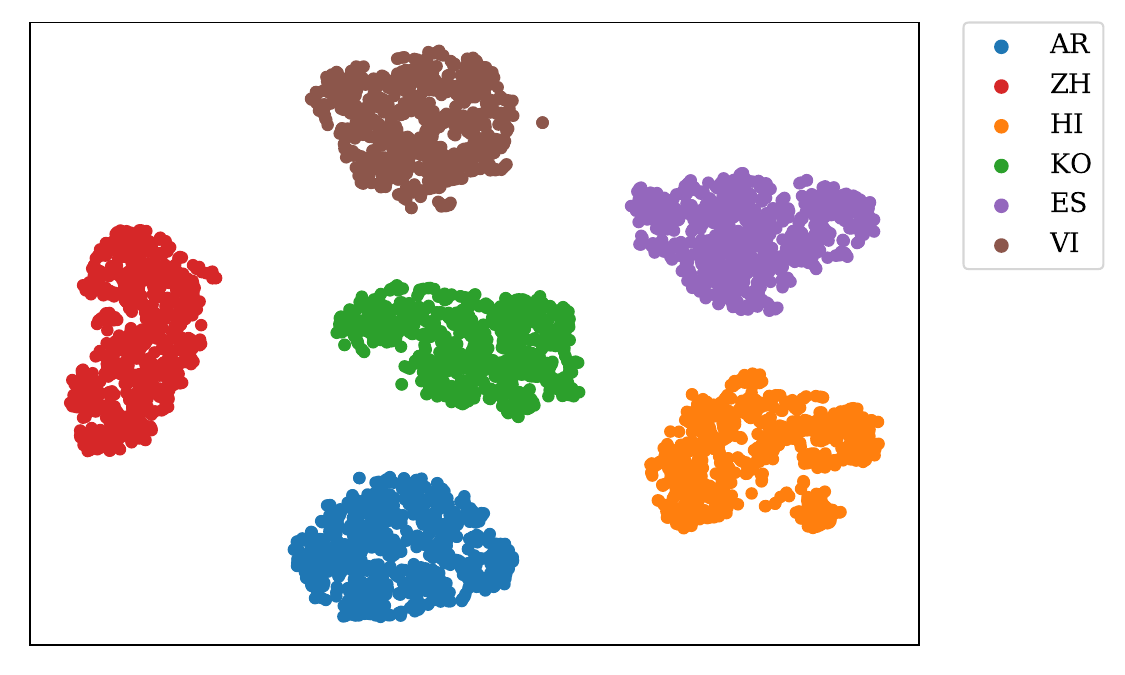}}
    \centerline{(c) SIGAM}
    \label{fig:c}
\end{minipage}
\caption{Visualizations of utterance level embeddings extracted from the ground truth accented speech by different models: (a) GST, (b) VAE, (c) SIGAM.}
\label{fig:accent_emb}
\end{figure*}

To better understand the performance differences on accent rendering, we visualize utterance level embeddings generated by the GST, VAE, and SIGAM from the ground truth accented speech to analyze the accent distribution. We randomly select 200 training utterances per speaker, resulting in 800 utterances per accent. The t-SNE \cite{van2008visualizing} visualizations are shown in Fig. \ref{fig:accent_emb}, where each data point represents an utterance level embedding vector. 
Fig. \ref{fig:accent_emb}(c) clearly illustrates that embeddings generated by the SIGAM 
cluster closely within the same accent while maintaining distinct separation between different accents, demonstrating the ability of the SIGAM to produce the accent-discriminative embedding $H_{G}$.
However, embeddings generated by the GST model in Fig. \ref{fig:accent_emb}(a) exhibit unclear boundaries between different accents, with only partial clustering within the same accent. 
In contrast, Fig. \ref{fig:accent_emb}(b) shows the weakest performance in accent modeling, since embeddings within the same accent lack recognizable clustering trends. 
This suggests that embeddings generated by the VAE model contain significantly less accent-discriminative information, which may explain the notably lowest AMOS scores of AM-VAE in the \emph{multi-speaker cross-accent} speech synthesis scenario.
As a result, the speech generated by AM-VAE may rely more heavily on the speaker embedding, potentially accounting for its highest SMOS scores. 

\subsubsection{Effectiveness of the SILAM with LAPM} 
To investigate the importance of the phoneme level SILAM with LAPM, we compare AM-SIMSAM with AM-SIGAM.
In Table \ref{tab:mcd_f0_dur_inherent}, we observe that AM-SIMSAM outperforms AM-SIGAM on MCD, $F0$ RMSE, $F0$ correlation, and FD across all accents, suggesting that the SILAM contributes to enhancing both speech quality and accent rendering. 
The accent XAB preference test results in Table \ref{tab:xab_accent_inherent_cross} show the significant preference for AM-SIMSAM over AM-SIGAM in both scenarios. This preference is particularly pronounced in the \emph{multi-speaker cross-accent} scenario, where AM-SIMSAM achieves a 126\% relative increase in accent preference compared to AM-SIGAM.
These findings strongly demonstrate the effectiveness of the SILAM on improving accent similarity in the generated speech by capturing fine-grained accent characteristics. 
Furthermore, they highlight the capability of the LAPM to predict local accent representations directly from the phoneme sequence.

\subsubsection{Effectiveness of speaker disentanglement}
\label{sec:ablation_study_speaker_disentanglement}

\begin{table}
\caption{Results of SECS and speaker XAB preference test for speaker similarity. NP denotes No Preference.}
\centering
\label{tab:xab_speaker_secs_cross}
\renewcommand\arraystretch{1}{
\scalebox{0.81}{
\begin{tabular}{p{0.8cm}<{\centering}|p{1.45cm}<{\centering}p{1.66cm}<{\centering}|p{1.45cm}<{\centering}p{1cm}<{\centering}p{1.66cm}<{\centering}} 
\hline
\specialrule{0em}{0pt}{2pt}
  &  \multicolumn{2}{c|}{SECS} & \multicolumn{3}{c}{Speaker XAB Preference (\%)} \\ 
\specialrule{0em}{2pt}{0pt}
\cline{2-6}
\specialrule{0em}{0pt}{2pt}
       & AM-MSAM    & AM-SIMSAM & AM-MSAM     &NP &  AM-SIMSAM    \\ 
\specialrule{0em}{2pt}{0pt}   
\hline
\specialrule{0em}{0pt}{2pt}    
  \multicolumn{6}{c}{\emph{Multi-speaker cross-accent}} \\ 
\specialrule{0em}{0pt}{2pt}
\hline
\specialrule{0em}{0pt}{2pt}
AR     & 0.825 & \textbf{0.851} & 34.17 & 10.83 & \textbf{55.00}             \\ 
HI     & 0.779 & \textbf{0.839} & 29.17 & 13.33 & \textbf{57.50}             \\
KO     & 0.835 & \textbf{0.855} & 32.50 & 17.50 & \textbf{50.00}             \\ 
ES     & 0.868 & \textbf{0.875} & 32.50 & 13.33 & \textbf{54.17}             \\ 
VI     & 0.819 & \textbf{0.855} & 35.00 & 8.33  & \textbf{56.67}             \\ 
\specialrule{0em}{2pt}{0pt}
\hline
\specialrule{0em}{2pt}{0pt}
AVG    & 0.825 & \textbf{0.855} & 32.67 & 12.66 & \textbf{54.67}             \\
\specialrule{0em}{2pt}{0pt}
\hline
\end{tabular}}}
\end{table}

We compare AM-SIMSAM with AM-MSAM in the \emph{multi-speaker cross-accent} speech synthesis scenario to evaluate the effectiveness of speaker disentanglement.
To ensure a fair comparison, both systems utilize reference speech during inference. 
In Table \ref{tab:xab_speaker_secs_cross}, the SECS results suggest that AM-SIMSAM achieves significantly higher speaker similarity than AM-MSAM. This is further supported by 
the speaker XAB preference test, which shows that AM-SIMSAM is significantly preferred over AM-MSAM on speaker similarity. 
These results demonstrate the effectiveness of speaker disentanglement on preserving target speaker identity.
In contrast, when MSAM extracts accent representations from the Mel-spectrogram of a source speaker, these representations are inherently entangled with source speaker identity, resulting in lower target speaker similarity in the generated speech.

\begin{figure*}[!t]
\centering
\begin{minipage}{.16\textwidth}
    \centerline{\includegraphics[width=3cm, height=2.25cm]{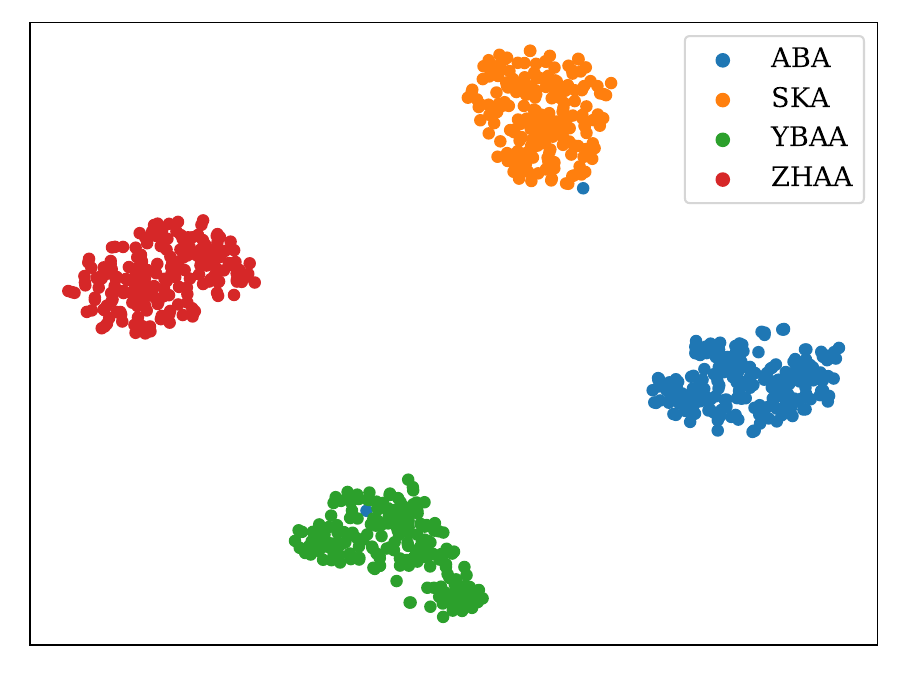}}
    \label{fig:a}
\end{minipage}
\begin{minipage}{.16\textwidth}
    \centerline{\includegraphics[width=3cm, height=2.25cm]{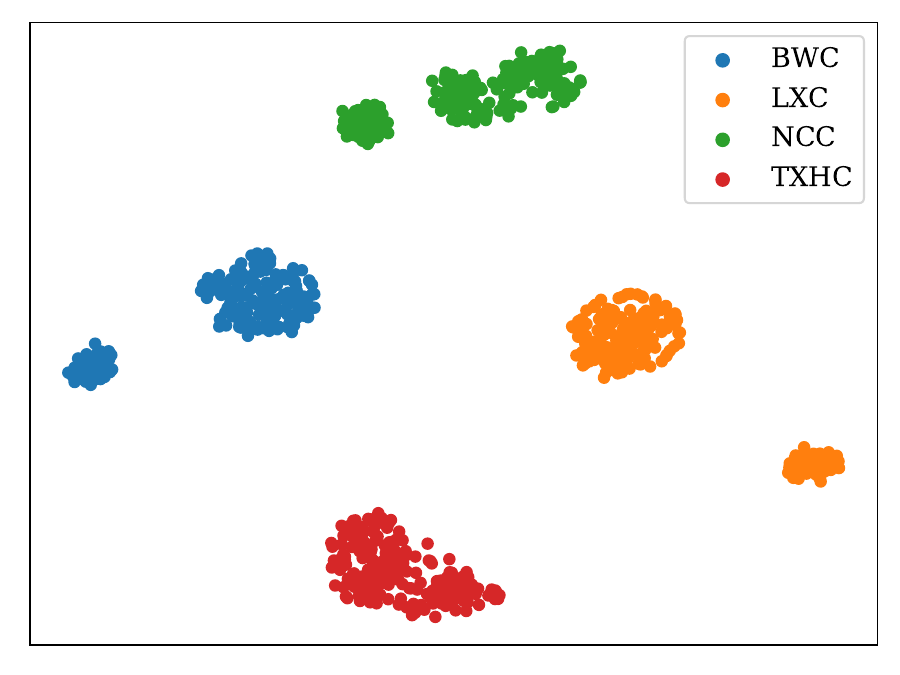}}
    \label{fig:b}
\end{minipage}
\begin{minipage}{.16\textwidth}
    \centerline{\includegraphics[width=3cm, height=2.25cm]{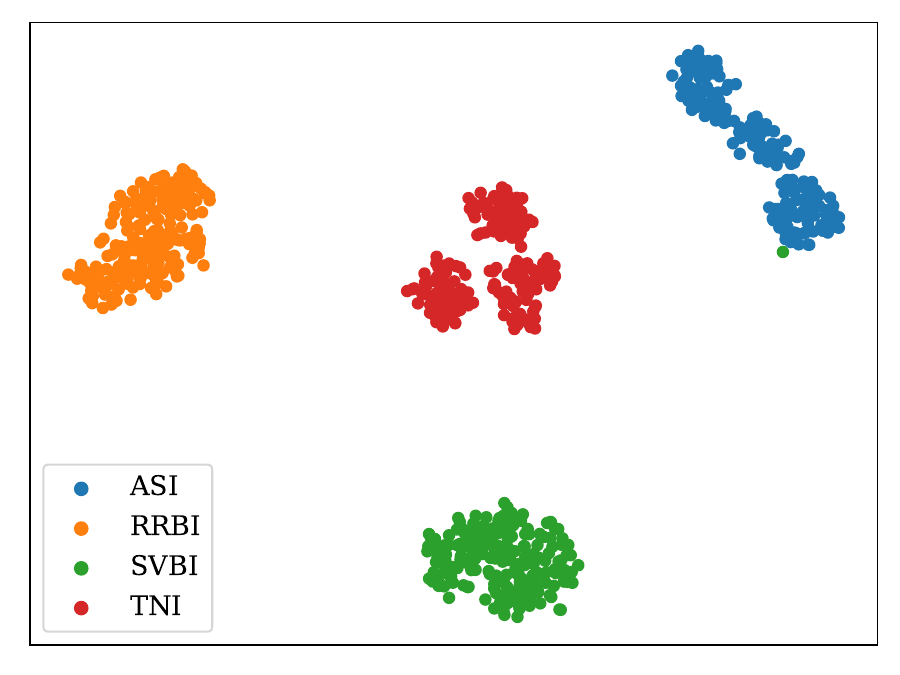}}
    \label{fig:c}
\end{minipage}
\begin{minipage}{.16\textwidth}
    \centerline{\includegraphics[width=3cm, height=2.25cm]{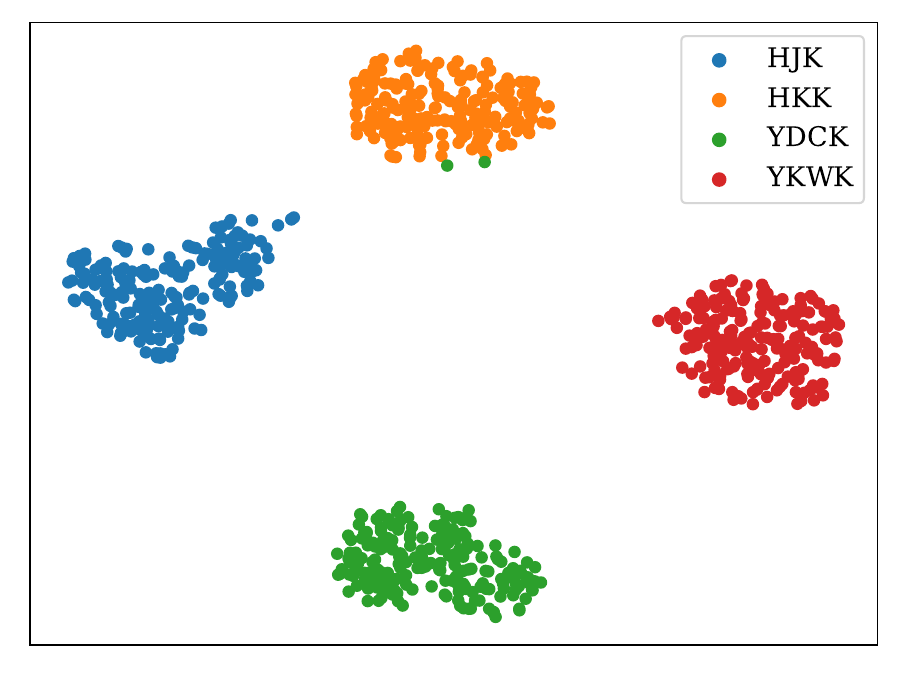}}
    \label{fig:c}
\end{minipage}
\begin{minipage}{.16\textwidth}
    \centerline{\includegraphics[width=3cm, height=2.25cm]{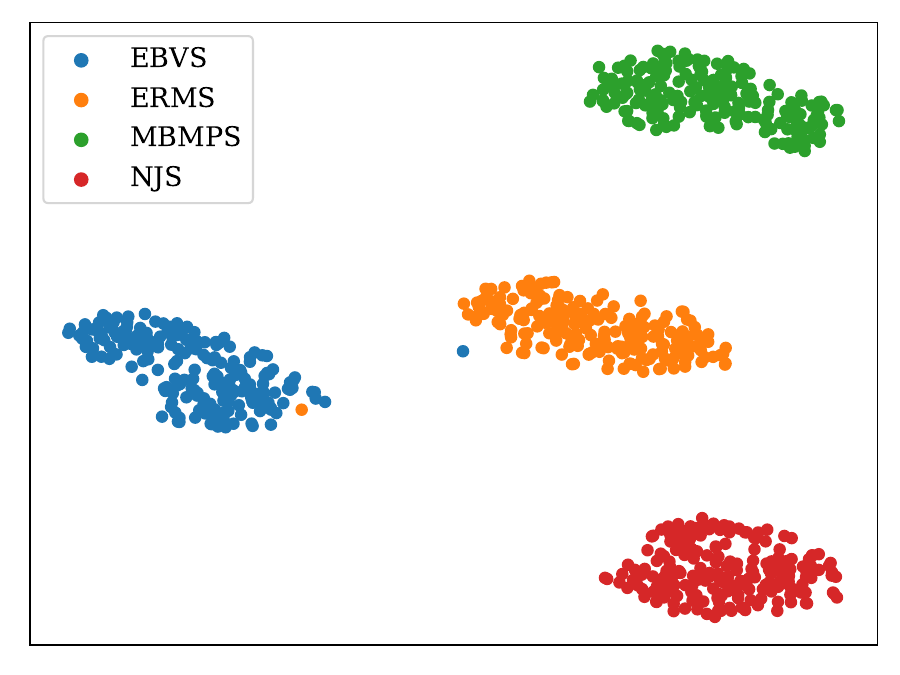}}
    \label{fig:c}
\end{minipage}
\begin{minipage}{.16\textwidth}
    \centerline{\includegraphics[width=3cm, height=2.25cm]{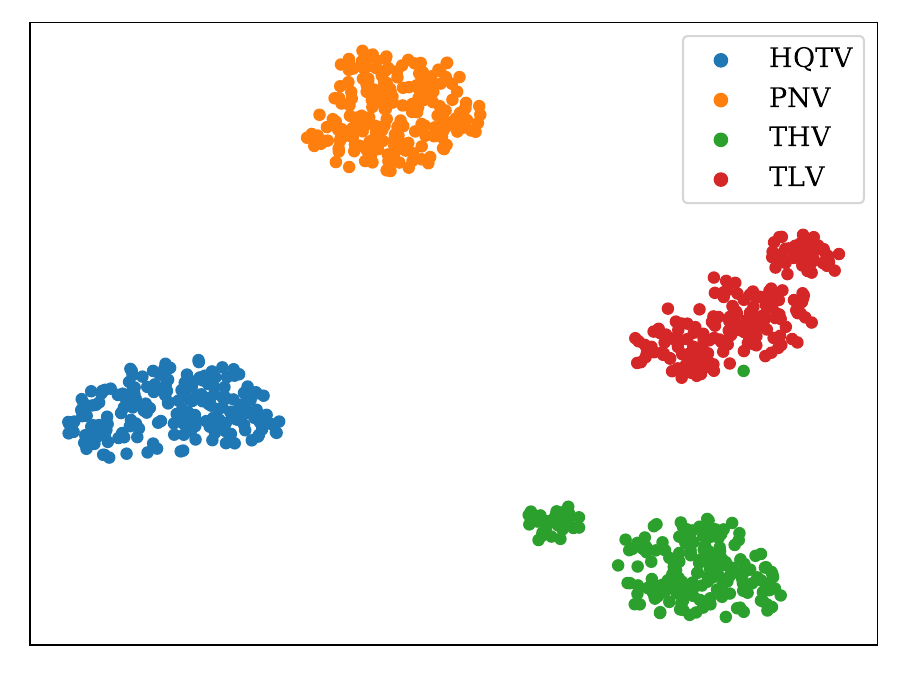}}
    \label{fig:c}
\end{minipage}
\quad
\begin{minipage}{.16\textwidth}
    \centerline{\includegraphics[width=3cm, height=2.25cm]{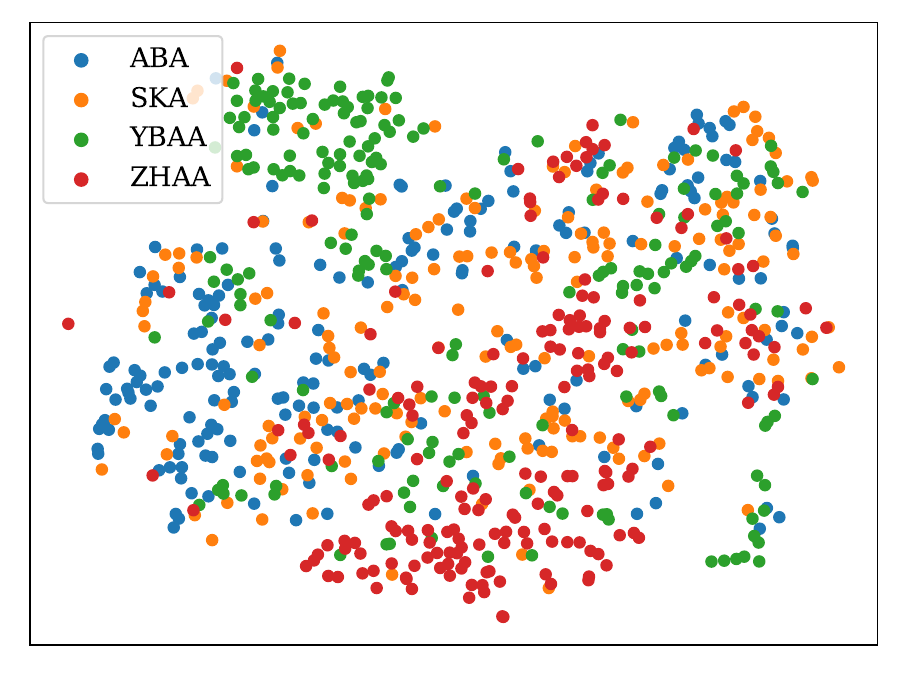}}
    \centerline{AR}
    \label{fig:a}
\end{minipage}
\begin{minipage}{.16\textwidth}
    \centerline{\includegraphics[width=3cm, height=2.25cm]{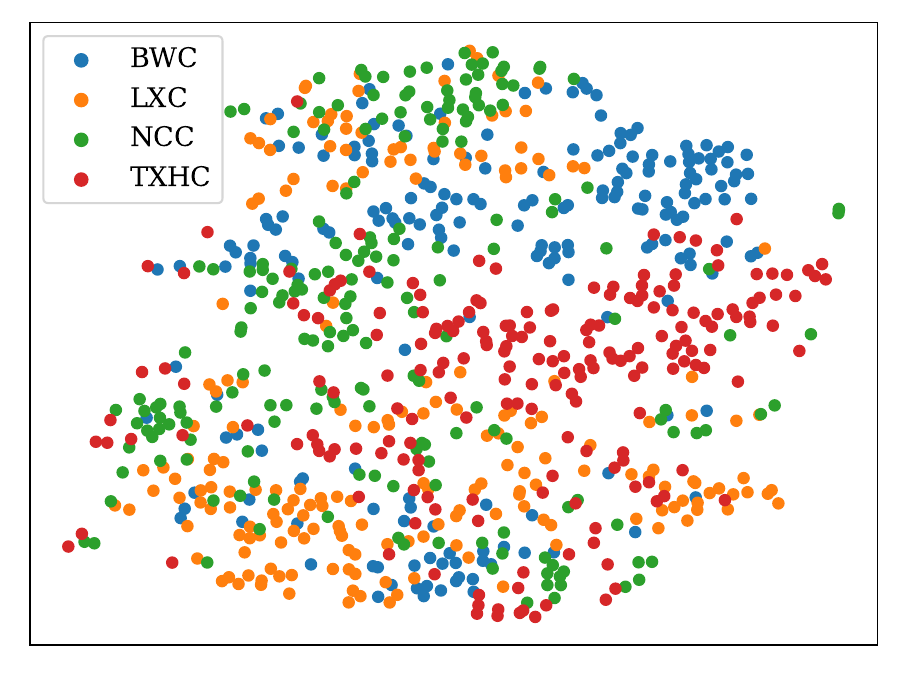}}
    \centerline{ZH}
    \label{fig:b}
\end{minipage}
\begin{minipage}{.16\textwidth}
    \centerline{\includegraphics[width=3cm, height=2.25cm]{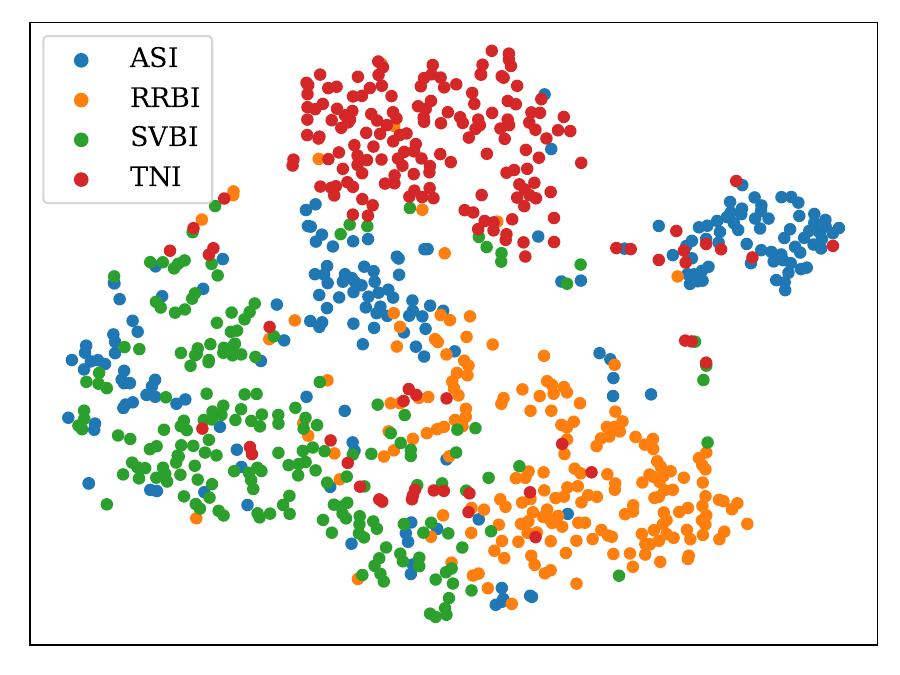}}
    \centerline{HI}
    \label{fig:c}
\end{minipage}
\begin{minipage}{.16\textwidth}
    \centerline{\includegraphics[width=3cm, height=2.25cm]{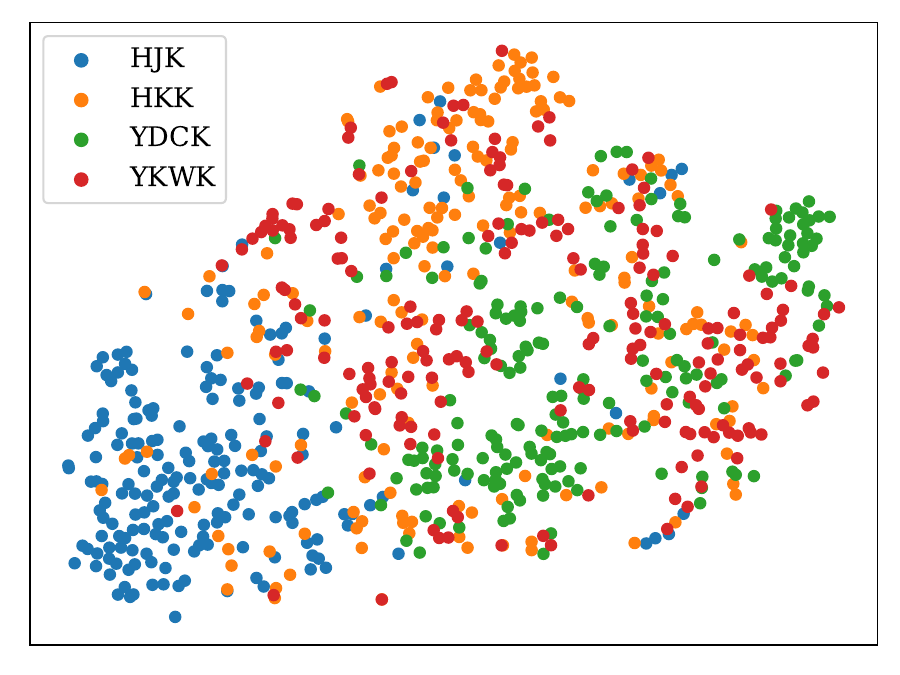}}
    \centerline{KO}
    \label{fig:c}
\end{minipage}
\begin{minipage}{.16\textwidth}
    \centerline{\includegraphics[width=3cm, height=2.25cm]{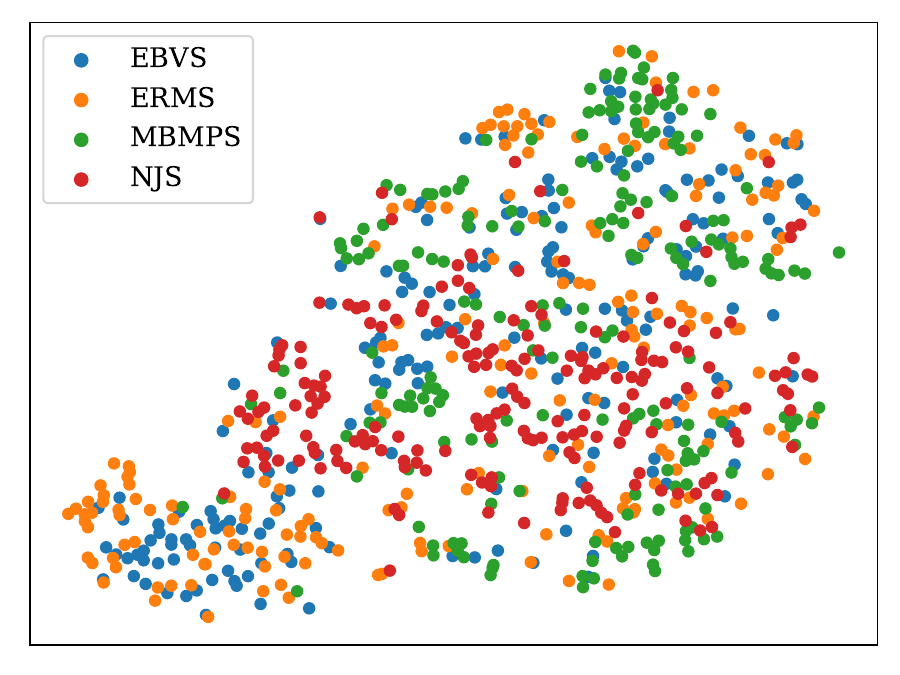}}
    \centerline{ES}
    \label{fig:c}
\end{minipage}
\begin{minipage}{.16\textwidth}
    \centerline{\includegraphics[width=3cm, height=2.25cm]{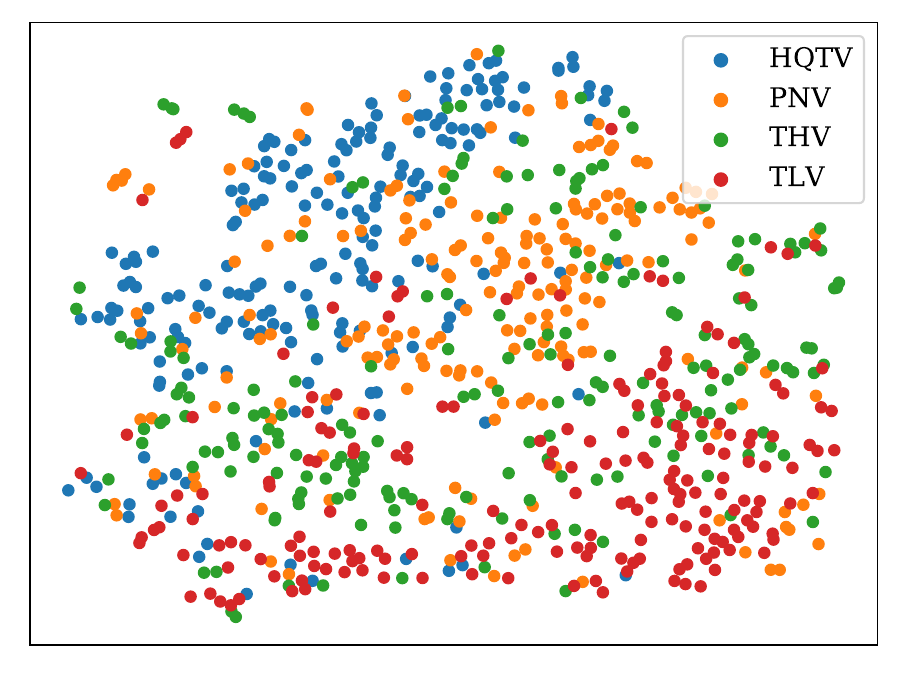}}
    \centerline{VI}
    \label{fig:c}
\end{minipage}
\caption{Visualizations of utterance level accent embeddings extracted from the ground truth accented speech by two models: GAM in the first row and SIGAM in the second row.}
\label{fig:accent_emb_ablation}
\end{figure*}

\begin{table}
\caption{Results of AECS for accent similarity.}
\centering
\label{tab:aecs}
\renewcommand\arraystretch{1}{
\begin{tabular}{p{1cm}<{\centering}|p{0.55cm}<{\centering}p{0.55cm}<{\centering}p{0.55cm}<{\centering}p{0.55cm}<{\centering}p{0.55cm}<{\centering}p{0.55cm}<{\centering}p{0.55cm}<{\centering}} 
\hline
\specialrule{0em}{0pt}{2pt}
& AR &ZH&HI&KO&ES&VI&AVG\\
\specialrule{0em}{0pt}{2pt}
\hline
\specialrule{0em}{0pt}{2pt}
GAM & 0.704 & 0.753 & 0.680 & 0.772 & 0.855 & 0.775 & 0.757 \\ 
SIGAM & \textbf{0.950} & \textbf{0.911} & \textbf{0.933} & \textbf{0.940} &\textbf{0.952} & \textbf{0.947}&\textbf{0.939} \\
\specialrule{0em}{0pt}{2pt}
\hline
\end{tabular}}
\end{table}

Furthermore, we visualize utterance level accent embeddings of each accent category generated by the GAM and SIGAM using t-SNE \cite{van2008visualizing}, as shown in Fig. \ref{fig:accent_emb_ablation}. 
In the first row, it is obvious that embeddings generated by the GAM exhibit clear boundaries corresponding to speaker identity within each accent, suggesting that these embeddings are speaker-dependent.  
In contrast, the second row illustrates embeddings generated by the SIGAM clustering more closely within each accent, 
which demonstrates that the SIGAM models accents in a speaker-independent manner.
Accent similarity of these embeddings is measured using AECS, as reported in Table \ref{tab:aecs}. The results indicate that the SIGAM achieves significantly higher accent similarity for each accent compared to the GAM, further validating that speaker information is effectively disentangled in the SIGAM.

\section{Conclusion and Future Work}
\label{sec:conclusion}
We propose a multi-speaker multi-accent TTS framework with the multi-scale accent modeling and disentangling approach at both global and local scales. 
The speaker-independent global accent model (SIGAM) and speaker-independent local accent model (SILAM) are introduced to comprehensively capture accent characteristics, including the overall fluctuations of an utterance and fine-grained variations across phonemes. 
Speaker disentanglement is further performed to enable speaker-independent accent modeling for flexible multi-speaker multi-accent speech synthesis.
We validate the effectiveness of the SIGAM and SILAM on accent rendering, with the local accent prediction model (LAPM) complementing the SILAM in the practical inference stage.
The speaker disentanglement also proves effective on preserving target speaker identity.
Experimental results demonstrate that our proposed system improves both speech quality and accent rendering while maintaining acceptable speaker similarity for multi-speaker multi-accent speech synthesis.
However, several limitations still remain, involving possible directions for future work.

While our approach enhances accent rendering in the generated speech, we acknowledge the trade-offs between accent rendering and speaker similarity in the \emph{multi-speaker cross-accent} speech synthesis scenario. Specifically, our proposed framework exhibits reduced performance on speaker similarity compared to AM-VAE in this scenario, highlighting the challenge of balancing accent rendering and speaker similarity.
Exploring an ideal strategy for generating multi-speaker multi-accent speech with both accurate accent rendering and high speaker similarity would be a valuable direction for future work.

In this study, we evaluate the performance of our system on generating multi-accent speech for seen speakers. However, the performance for unseen speakers remains to be explored. Another valuable future work is to extend our system to zero-shot multi-speaker scenarios, which could be more beneficial and practical for read-world applications.

Note that we use Tacotron 2 as the AM in this study. Our research focuses on investigating techniques that can be applied to various TTS architectures for multi-speaker multi-accent speech synthesis, rather than being limited to a specific TTS model. Tacotron 2 serves as an illustrative example to demonstrate the effectiveness of our approach. 
In future work, we plan to extend our approach to other TTS models.

\bibliographystyle{IEEEtran}
\bibliography{ref}

\vfill

\end{document}